\documentclass[conference]{IEEEtran}
\usepackage{verbatim}

\usepackage{leftidx}
\usepackage{hyperref}
\usepackage{cuted}
\usepackage{multicol}

\ifCLASSINFOpdf
\else
\fi
\usepackage[caption=false,font=footnotesize]{subfig}
\newcounter{MYtempeqncnt}
\newcounter{nexttempeqncnt}

\usepackage{stfloats}

\usepackage{caption}

\usepackage{tabu}

\usepackage[cmex10]{amsmath}
\usepackage{amssymb}
\usepackage{graphicx}
\newtheorem{theorem}{Theorem}

\newtheorem{lemma}{Lemma}

\renewcommand\footnotemark{}

\newcommand\blfootnote[1]{%
\begingroup
\renewcommand\thefootnote{}\footnote{#1}%
\addtocounter{footnote}{-1}%
\endgroup
}
\usepackage[norule,symbol,perpage]{footmisc}
\begin{document}
\date{}
\title{Interference Mitigation Using Asynchronous Transmission and Sampling Diversity}
\author{ \normalsize Mehdi Ganji and Hamid Jafarkhani }
\maketitle

\begin{abstract}
In this paper, we show that by investigating inherent time delays between different users in a multiuser scenario, we are able to cancel interference more efficiently. Time asynchrony provides another tool to cancel interference which results in preserving other resources like frequency, time and code. Therefore, we can save the invaluable resource of frequency band and also increase spectral efficiency. A sampling method is presented which results in independent noise samples and obviates the need for the complex process of noise whitening. By taking advantage of this sampling method and its unique structure, we implement maximum-likelihood sequence detection which outperforms synchronous maximum-likelihood detection. We also present successive interference cancellation with hard decision passing which gives rise to a novel forward-backward belief propagation method. Next, the performance of zero forcing detection is analyzed. Simulation results are also presented to verify our analysis. \\

\end{abstract}

\vspace*{-0.5em}
\section{Introduction}
\footnotetext{\llap{\textsuperscript{}}This work was supported in part by the NSF Award CCF--–1526780. The authors are with the Center for Pervasive Communications and
Computing, University of California, Irvine, CA 92697-2625 USA (e-mail:
\{mganji, hamidj\}@uci.edu).}
There are many applications where multiple users share a common channel to transmit data to a receiver. Numerous examples of multiaccess communication include uplink transmission of a single cell in a cellular system, a group of twisted-pair copper subscriber lines transmitting data to the same switching office, multiple ground stations communicating with a satellite and interactive cable television networks. The key challenge in multiuser transmissions or multiple access channels is Interuser Interference. Over several decades, many methods have been introduced to address this problem \cite{honig2009advances}, \cite{verdu1998multiuser}. Most of these methods are based on assigning orthogonal dimensions to different users to be able to separate them and prevent interference. For example, time division multiple access (TDMA) protocols allocate different time slots to different users to mitigate interference. The same concept can be applied by partitioning the frequency spectrum among different users, which is called frequency division multiple access (FDMA). Code division multiple access is another scheme used to surpass interuser interference in which users are multiplexed by distinct codes rather than by orthogonal frequency bands, or by orthogonal time slots \cite{moshavi1996multi}. More recently, multiple receive antennas are utilized at the receive side to take advantage of the spatial domain in order to cancel interference \cite{naguib1998applications}, \cite{kazemitabar2008multiuser}.

In this paper, we investigate the timing mismatch between users as an additional resource to address the problem of interuser interference. By exploiting time delays between users and employing an appropriate sampling method, we design detection methods which not only cancel the interference effectively, but also outperform the synchronous ones. When timing mismatch is used to cancel the interuser interference, resources like frequency spectrum, time and receive antenna can be employed to improve the performance. There are other examples in the literature in which asynchronous transmission outperforms synchronous transmission. For example, by using timing delays between users, zero forcing (ZF) detection can be performed with one receive antenna and additional receive antennas can be used to gain diversity \cite{shao2007performance}, \cite{das2011mimo}. In \cite{shao2007performance}, the authors proposed a ZF receiver in MIMO setting which takes advantage of timing mismatch between data streams and provides full diversity of $M$, where $M$ is the number of receive antennas. However, a crucial impairment of their receiver design is addressed in \cite{das2011mimo}. The design of asynchronous differential decoding methods which outperform their synchronous counterparts is discussed in \cite{poorkasmaei2015asynchronous}, \cite{avendi2015differential}. In this paper, we present sampling diversity and provide several decoders to gain advantages from asynchronous transmission. We analytically prove that our ZF method provides full diversity and we study its asymptotic performance for large number of receive antennas.

\vspace*{-0.5em}
\section{System Model}
\subsection{General Settings}
We consider a system with K users, transmitting data to a common receiver simultaneously, which can have one receive antenna or multiple ones. Due to different physical locations of users, their signal is received with various time delays. It is assumed that each data stream is received with an arbitrary delay smaller than the symbol interval and only the receiver knows the time delays. The signal transmitted from User k is described by:
\begin{eqnarray}
&s_k(t)=\sum_{i=1}^{N}{b_k(i)p(t-(i-1)T_s)} \ \ 
\end{eqnarray}
where $T_s$ is the symbol length and $p(.)$ is the pulse-shaping filter with non-zero duration of $T$. Also, $N$ is the frame length and $b_k(i)$ is the transmitted symbol by User k in the $i$th time slot. The transmitted signals are received with a relative delay of $\tau_{k}$ and a channel path gain of $h_{k}$.
Then, the received signal can be represented by:
\begin{eqnarray}
&y(t)=\sum_{k=1}^{K}{h_{k}s_k(t-\tau_{k})}+n(t) 
\end{eqnarray}
where $K$ is the number of users and $n(t)$ is the white noise with variance of $\sigma^2$. Without loss of generality, we assume that $0=\tau_{1}<\tau_{2}<\dots<\tau_{K}<T$.

\subsection{Output Samples}
\begin{figure}[!h]
\centering
\subfloat[\label{asynch2}]{%
\includegraphics[width=3in]{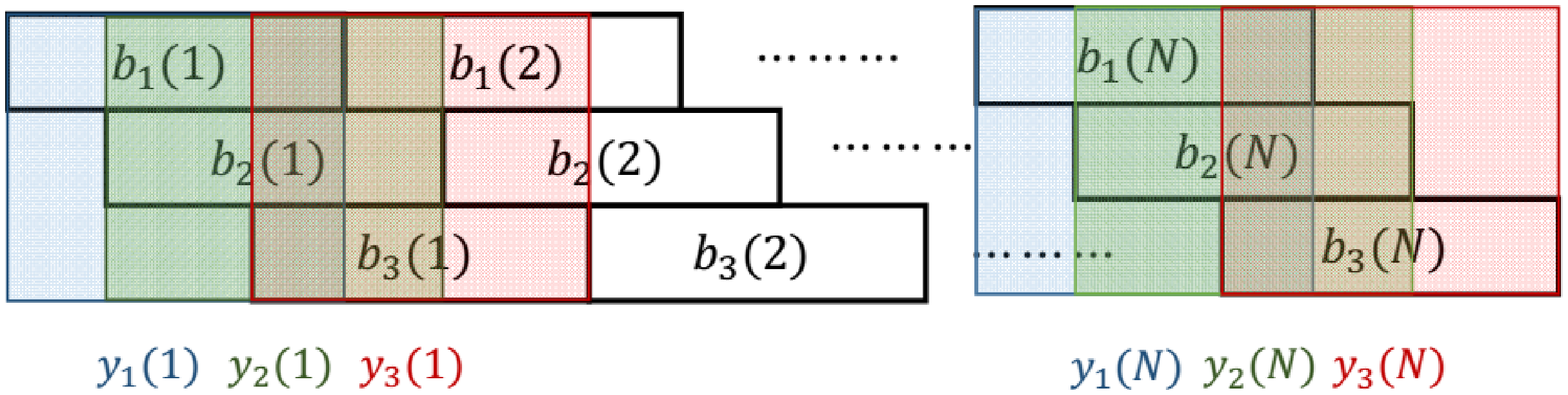}
}\par
\subfloat[\label{asynch}]{%
\includegraphics[width=3in]{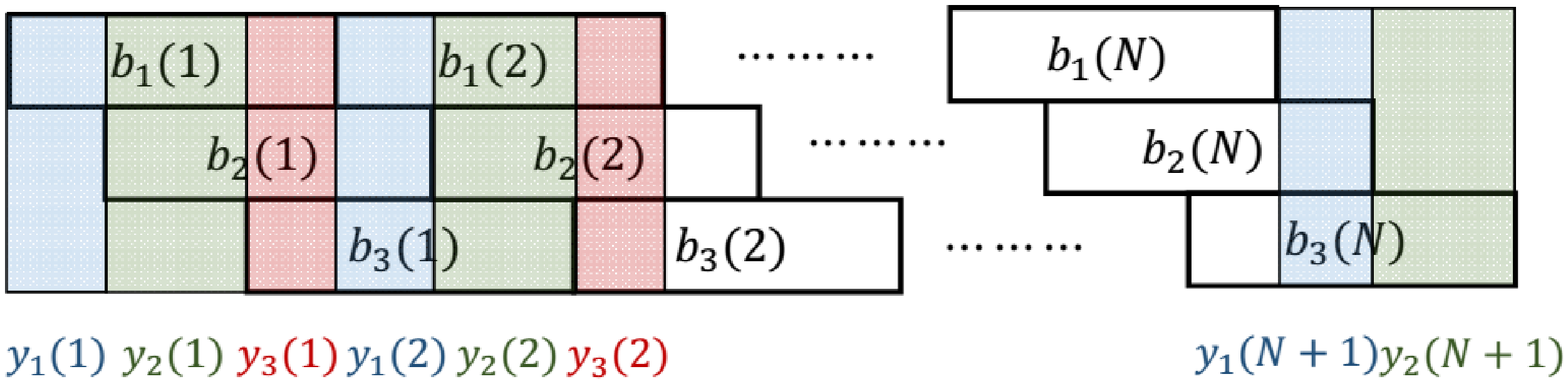}
} 
\caption{Sampling methods}
\label{TS}
\end{figure}

The output of the matched filter at each receiver antenna can be sampled at different sampling times associated with different users as shown in Fig. \ref{asynch2}. These sets of samples provide sufficient statistics for decoding transmitted symbols \cite{verdu}. We can break down the integrals corresponding to the sampling in Fig. \ref{asynch2} to define a new sampling method as shown in Fig. \ref{asynch}. The corresponding output samples are written in Eq. (\ref{eq:sampling1_1}) where $\tau_{k+1}$ is an auxiliary variable equal to T. \\
\begin{figure*}[!b]
\normalsize
\setcounter{MYtempeqncnt}{\value{equation}}
\setcounter{equation}{8}
\hrulefill
\begin{flalign} 
\nonumber
y_{l}(j)&=\int_{\tau_{l}+(j-1)T_s}^{\tau_{(l+1)}+(j-1)T_s}{\sum_{k=1}^{K}{\sum_{i=1}^{N}{b_k(i)p(t-(j-1)T_s-\tau_{l})p(t-(i-1)T_s-\tau_{k})}h_{k}}dt}\\
\label{eq:sampling1_1}
&+\int_{\tau_{l}+(j-1)T_s}^{\tau_{(l+1)}+(j-1)T_s}{n(t)p(t-(j-1)T_s-\tau_{l})dt} \ \ \ \ \ \ 1\leq l\leq K,\ 1\leq j \leq N+1\\
\label{eq:sampling1_2}
u_{ji}(l,k)&=\int_{\tau_{l}+(j-1)T_s}^{\tau_{(l+1)}+(j-1)T_s}{p(t-(j-1)T_s-\tau_{l})p(t-(i-1)T_s-\tau_{k})dt}\\
\label{eq:sampling1_3}
v_{l}(j)&=\int_{\tau_{l}+(j-1)T_s}^{\tau_{(l+1)}+(j-1)T_s}{n(t)p(t-(j-1)T_s-\tau_{l})dt}
\end{flalign}
\setcounter{nexttempeqncnt}{\value{equation}}
\setcounter{equation}{\value{MYtempeqncnt}}
\end{figure*}
By defining intermediate variables $u_{ji}(l,k)$ and noise samples $v_{l}(j)$ in Eqs. (\ref{eq:sampling1_2}) and (\ref{eq:sampling1_3}), at the bottom of the page, we can write output samples in a more compact way:
\begin{flalign} 
y_{l}(j)&=\sum_{k=1}^{K}{\sum_{i=1}^{N}{b_k(i) u_{ji}(l,k) h_{k}}}+v_{l}(j)\\
\nonumber
j&=1,\dots,N+1 \ \ \text{sampling time index} \\
\nonumber
l&=1,\dots,K \ \ \text{index of matched user}
\end{flalign} 
Defining $\boldsymbol{y(j)}=[y_1(j),y_2(j),\dots, y_K(j)]^T$ and $\boldsymbol{b(i)}=[b_1(i),b_2(i),\dots, b_K(i)]^T$, then, $\boldsymbol{y(j)}$ for different values of $j$ can be written as:
\begin{flalign}
\boldsymbol{y(j)}=\sum_{i=1}^{N}{\boldsymbol{U_{ji}}\boldsymbol{h}\boldsymbol{b(i)}}+\boldsymbol{v(j)} \ \ 1\leq j \leq N+1
\end{flalign}
where $\boldsymbol{h}=\text{diag}[h_1,h_2,\dots,h_K]$, $\boldsymbol{v(j)}=[v_1(j),v_2(j),\dots,v_K(j)]^T$ and $\boldsymbol{U_{ji}}$ is a $K\times K$ matrix whose elements are defined as $\boldsymbol{U_{ji}}(l,k)=u_{ji}(l,k)$. The next step is to put all vectors of $\boldsymbol{y(j)}$ together and define $\boldsymbol{y}$ as $[\boldsymbol{y(1)},\boldsymbol{y(2)},\dots,\boldsymbol{y(N+1)}]^T$. Then, $\boldsymbol{y}$ can be written as:
\begin{flalign}\label{eq:sampling1_original} 
\nonumber
&\boldsymbol{y} =\\
\nonumber
&\left(\begin{smallmatrix}
\boldsymbol{U_{11}}&\boldsymbol{U_{12}}& \boldsymbol{U_{13}}&\dots & \boldsymbol{U_{1N}}\\ 
\boldsymbol{U_{21}} & \boldsymbol{U_{11}}&\boldsymbol{U_{12}}&\dots &\boldsymbol{U_{1(N-1)}}\\ 
\vdots&\ddots & \ddots &\ddots &\vdots\\ 
\boldsymbol{U_{(N-1)1}}&\dots&\boldsymbol{U_{21}}&\boldsymbol{U_{11}}&\boldsymbol{U_{12}}\\ 
\boldsymbol{U_{N1}}&\dots&\boldsymbol{U_{31}}&\boldsymbol{U_{21}}&\boldsymbol{U_{11}}\\
\boldsymbol{U_{(N+1)1}}&\dots&\boldsymbol{U_{41}}&\boldsymbol{U_{31}}&\boldsymbol{U_{21}}
\end{smallmatrix}\right)
\left(
\begin{smallmatrix}
\boldsymbol{h}&\boldsymbol{0}& \boldsymbol{0}&\dots & \boldsymbol{0}\\ 
\boldsymbol{0} & \boldsymbol{h}&\boldsymbol{0}& \dots & \boldsymbol{0}\\ 
\vdots&\ddots & \ddots &\ddots & \vdots\\ 
\boldsymbol{0}&\dots& \boldsymbol{0}& \boldsymbol{h} &\boldsymbol{0} \\ 
\boldsymbol{0}&\dots&\boldsymbol{0}&\boldsymbol{0}&\boldsymbol{h}
\end{smallmatrix}
\right)\left(\begin{smallmatrix}
\boldsymbol{b(1)}\\ 
\boldsymbol{b(2)}\\ 
\vdots\\ 
\boldsymbol{b(N)}
\end{smallmatrix}\right)+\boldsymbol{v}\\
&=\boldsymbol{U}\boldsymbol{H} \boldsymbol{b}+\boldsymbol{v}
\end{flalign}
Block Toeplitz structure of $\boldsymbol{U}$ originates from the fact that $u_{(j+m)(i+m)}(l,k)=u_{ji}(l,k)$. This can be verified by a change of variable in Eq. (\ref{eq:sampling1_2}). Based on the relation between $T$ and $T_s$, different numbers of adjacent symbols interfere with each other. For example, for rectangular pulse shapes, i.e., $T=T_s$, at each instant only current and previous symbols cause interference. In other words, only $\boldsymbol{U_{11}}$ and $\boldsymbol{U_{21}}$ are nonzero. Without loss of generality, we assume that $T=1$, therefore $\boldsymbol{U_{11}}$ and $\boldsymbol{U_{21}}$ are defined as follows:
\begin{flalign}
&\boldsymbol{U_{11}}=
\left(\begin{smallmatrix}
\tau_2-\tau_1&0& \dots & 0\\ 
\tau_3-\tau_2 & \tau_3-\tau_2 & \dots & 0\\ 
\vdots&\vdots & \ddots & \vdots\\ 
\tau_{K}-\tau_{K-1}& \dots & \tau_{K}-\tau_{K-1} &0 \\ 
1-\tau_K &\dots &1-\tau_K&1-\tau_K 
\end{smallmatrix}\right)
\end{flalign}
\begin{flalign}
\boldsymbol{U_{21}}=
\left(\begin{smallmatrix}
0&\tau_2-\tau_1& \dots & \tau_2-\tau_1\\ 
0 & 0 & \dots & \tau_3-\tau_2\\ 
\vdots&\vdots & \ddots & \vdots\\ 
0& \dots & 0 &\tau_{K-1}-\tau_K \\ 
0 &\dots &0&0
\end{smallmatrix}\right)
\end{flalign}
Hence, for rectangular pulse shapes, the system model simplifies to:
\begin{flalign}\label{eq:sampling1_rect} 
\nonumber
&\boldsymbol{y} =\\
\nonumber
&\left(\begin{smallmatrix}
\boldsymbol{U_{11}}&\boldsymbol{0}& \boldsymbol{0}&\dots & \boldsymbol{0}\\ 
\boldsymbol{U_{21}} & \boldsymbol{U_{11}}&\boldsymbol{0}& \dots & \boldsymbol{0}\\ 
\vdots&\ddots & \ddots &\ddots & \vdots\\ 
\boldsymbol{0}&\dots& \boldsymbol{U_{21}}& \boldsymbol{U_{11}} &\boldsymbol{0} \\ 
\boldsymbol{0}&\dots&\boldsymbol{0}&\boldsymbol{U_{21}} &\boldsymbol{U_{11}} \\
\boldsymbol{0}&\dots&\boldsymbol{0}&\boldsymbol{0} &\boldsymbol{U_{21}}
\end{smallmatrix}\right)
\left(
\begin{smallmatrix}
\boldsymbol{h}&\boldsymbol{0}& \boldsymbol{0}&\dots & \boldsymbol{0}\\ 
\boldsymbol{0}& \boldsymbol{h}&\boldsymbol{0}& \dots & \boldsymbol{0}\\ 
\vdots&\ddots & \ddots &\ddots & \vdots\\ 
\boldsymbol{0}&\dots& \boldsymbol{0}& \boldsymbol{h} &\boldsymbol{0} \\ 
\boldsymbol{0}&\dots&\boldsymbol{0}&\boldsymbol{0}&\boldsymbol{h}
\end{smallmatrix}
\right)\left(\begin{smallmatrix}
\boldsymbol{b(1)}\\ 
\boldsymbol{b(2)}\\ 
\vdots\\ 
\boldsymbol{b(N)}
\end{smallmatrix}\right)+\boldsymbol{v}\\
&=\boldsymbol{UH}\boldsymbol{b}+\boldsymbol{v}
\end{flalign}
The important fact about this sampling method is that the covariance matrix of noise samples is diagonal. With a small abuse of notation, we denote $Diag(\boldsymbol{U_{11}})$ as a diagonal matrix including diagonal elements of $\boldsymbol{U_{11}}$. Then, it can be shown that $E[\boldsymbol{v}\boldsymbol{v}^H]$ is equal to $\sigma^2(\boldsymbol{I_N} \otimes Diag(\textbf{U}_\textbf{{11}}))$, where $\boldsymbol{I_N}$ is an $N \times N$ identity matrix and $(\otimes)$ is Kronecker product.\\

Since the statistically sufficient samples in Fig. \ref{asynch2} can be created from samples in Fig. \ref{asynch}, the samples in Fig. \ref{asynch}, i.e. Eq. (\ref{eq:sampling1_rect}), are sufficient statistics too. Both of these sampling methods introduce intentional intersymbol interference (ISI) and impose memory on the system; however, they have some differences: 
\begin{enumerate}[\IEEEsetlabelwidth{12)}]
\item Sampling intervals in Fig. \ref{asynch} are smaller and need faster sampler.
\item Since sampling intervals are disjoint in Fig. \ref{asynch}, noise samples are independent. However, due to sampling overlap, the noise samples in Fig. \ref{asynch2} are correlated.
\item The sampling in Fig. \ref{asynch} results in an overdetermined system, while the number of output samples in Fig. \ref{asynch2} is equal to the number of input symbols.
\end{enumerate}

For rectangular pulse shape, the input-output relationship of the sampling method in Fig. \ref{asynch2} is:
\begin{flalign}\label{eq:sampling2_2} 
\nonumber
&\boldsymbol{y} =\\
\nonumber
&\left(\begin{smallmatrix}
\boldsymbol{R_{11}}&\boldsymbol{R_{12}}& \boldsymbol{0}&\dots & \boldsymbol{0}\\ 
\boldsymbol{R_{21}} & \boldsymbol{R_{11}}& \boldsymbol{R_{12}}& \dots & \boldsymbol{0}\\ 
\vdots&\ddots & \ddots &\ddots & \vdots\\ 
\boldsymbol{0}&\dots& \boldsymbol{R_{21}}& \boldsymbol{R_{11}} &\boldsymbol{R_{12}} \\ 
\boldsymbol{0}&\dots&\boldsymbol{0}&\boldsymbol{R_{21}} &\boldsymbol{R_{11}} 
\end{smallmatrix}\right)
\left(
\begin{smallmatrix}
\boldsymbol{h}&\boldsymbol{0}& \boldsymbol{0}&\dots & \boldsymbol{0}\\ 
\boldsymbol{0} & \boldsymbol{h}&\boldsymbol{0}& \dots & \boldsymbol{0}\\ 
\vdots&\ddots & \ddots &\ddots & \vdots\\ 
\boldsymbol{0}&\dots& \boldsymbol{0}& \boldsymbol{h} &\boldsymbol{0} \\ 
\boldsymbol{0}&\dots&\boldsymbol{0}&\boldsymbol{0} &\boldsymbol{h}
\end{smallmatrix}
\right)\left(\begin{smallmatrix}
\boldsymbol{b(1)}\\ 
\boldsymbol{b(2)}\\ 
\vdots\\ 
\boldsymbol{b(N)}
\end{smallmatrix}\right)+\boldsymbol{n}\\
\setcounter{equation}{\value{nexttempeqncnt}}
&=\boldsymbol{R H}\boldsymbol{ b}+\boldsymbol{n}
\end{flalign}
where $\boldsymbol{R}_{\boldsymbol{11}}$,$\boldsymbol{R}_{\boldsymbol{21}}$ and $\boldsymbol{R}_{\boldsymbol{12}}$ are defined as:

\begin{flalign}
\nonumber
&\boldsymbol{R}_{\boldsymbol{11}}=\\
&\left(\begin{smallmatrix}
1& 1-(\tau_{2}-\tau_{1})&\dots &1-(\tau_{K} - \tau_{1})\\ 
1-(\tau_{2}-\tau_{1})& 1&\dots & 1-(\tau_{K} - \tau_{2})\\ 
\vdots& \ddots & \ddots&\vdots\\ 
1-(\tau_{K-1}-\tau_{1})& \dots & 1&1-(\tau_{K} - \tau_{K-1})\\ 
1-(\tau_{K} - \tau_{1})& \dots &1-(\tau_{K} - \tau_{K-1})&1
\end{smallmatrix}\right)
\end{flalign}

\begin{flalign}\label{eq:R12} 
&\boldsymbol{R}_{\boldsymbol{12}}=(\boldsymbol{R}_{\boldsymbol{21}})^T=
\left(\begin{smallmatrix}
0&0& \dots & 0&0\\ 
\tau_{2}-\tau_{1} & 0& \dots &0& 0\\ 
\vdots&\ddots &\ddots& \vdots & \vdots\\ 
\tau_{(K-1)} - \tau_{1}&\tau_{(K-1)} - \tau_{2}& \dots &0&0 \\ 
\tau_{K} - \tau_{1}&\tau_{K} - \tau_{2}&\dots &\tau_{K} - \tau_{K-1}&0 
\end{smallmatrix}\right)
\end{flalign}
Because of intersection between sampling intervals, noise samples are correlated and noise whitening procedure needs to be performed before symbol detection. Noise whitening involves Cholesky decomposition and matrix inversion which increases complexity of receiver. 

\section{Receiver Design}\label{section:receiver}
In this section we introduce different detection methods which take advantage of distinct features of the sampling method shown in Fig. \ref{asynch}. One of these features is converting a memoryless system into a system with memory and independent noise samples. This enables us to implement the Viterbi algorithm based on samples in Eq. (\ref{eq:sampling1_rect}). The other feature is that this sampling method provides extra output samples which can be used to improve detection methods. For example, these extra samples make it possible to carry out successive interference cancellation (SIC) backward and forward. Also, by means of introduced ISI, zero forcing detection can be performed even with one receive antenna, which is impossible in synchronous multiuser transmission. In what follows, we will show how asynchronous multiuser transmission can outperform synchronous multiuser transmission. 

\subsection{Maximum-Likelihood Sequence Detection (MLSD)}
Due to inherent memory in the system that results from time delays, we can use the maximum-likelihood sequence detection method implemented by the Viterbi algorithm. The objective of maximum-likelihood sequence detector is to find the input sequence that maximizes the conditional probability, or the likelihood of the given output sequence. Exhaustive search over $2^{NK}$ different input sequences is an obvious choice, but it is impractical even for a moderate number of $K$ and $N$. Fortunately, using the Viterbi algorithm, MLSD can be implemented by complexity order of $2^K$ \cite{proakis2001intersymbol}. For using the Viterbi algorithm, the likelihood metric should be additive and the noise samples should be independent. Therefore, the sampling method in Fig. \ref{asynch} is the best fit for implementing the Viterbi algorithm. It reduces the complexity by avoiding the noise whitening procedure which involves Cholesky Decomposition and matrix inversion. We show that by using this sampling method, we can outperform synchronous ML detection with the same complexity order of $2^K$.\\
Based on the recursive relation between input and output which is described as:
\begin{flalign}
\nonumber
\boldsymbol{y(j)}=\boldsymbol{U_{11}hb(j)}-\boldsymbol{U_{21}hb(j-1)}+\boldsymbol{v(j)} \ \ 2\leq j \leq N
\end{flalign}
the trellis diagram of the system includes $A^K$ states with $A^K$ outgoing paths to the next states, and $A^K$ incoming paths from previous states, where $A$ is the size of the transmitted modulation. To calculate the metric for each path, we need to calculate the likelihood function as follows:
\begin{flalign}
\nonumber
&Pr(\boldsymbol{y(j)}|\boldsymbol{b(j)},\boldsymbol{b(j-1)})=\\
\nonumber
&Pr(\boldsymbol{v(j)}=\boldsymbol{y(j)}-\boldsymbol{U_{11}hb(j)}-\boldsymbol{U_{21}hb(j-1)})=\\
\nonumber
&\frac{1}{\sqrt{(2\pi)^K|\boldsymbol{\Sigma|}}}\exp{(-\frac{1}{2}\boldsymbol{s_j}^H\boldsymbol{\Sigma}^{-1}\boldsymbol{s_j})}
\end{flalign}
where $\boldsymbol{s_j}=\boldsymbol{y(j)}-\boldsymbol{U_{11}hb(j)}-\boldsymbol{U_{21}hb(j-1)}$ and $\boldsymbol{\Sigma}=E[\boldsymbol{v(j)}\boldsymbol{v(j)}^H]$. By discarding common terms and simple calculations, the metric for each path can be defined as $\sum_{i=1}^{K}{\frac{|\boldsymbol{s_j}(i)|^2}{\sigma^2\boldsymbol{U_{11}}(i,i)}}$. After calculating the path metrics, the final goal is to find the surviving path and trace it back to decode the transmitted symbols. The simulation result for this algorithm and its comparison with the synchronous ML detection is presented in Section \ref{section:Simulation Results}.

\subsection{Successive Interference Cancellation with Hard Decision Passing} \label{section:SIC}
Despite the excellent performance provided by MLSD, its complexity grows exponentially with the number of users, which might be prohibitive in some practical scenarios. Successive interference cancellation (SIC) detection that takes a serial approach to cancel interference can be used to reduce complexity. Using the sampling method in Fig. \ref{asynch}, this serial approach can be either a forward SIC initiated from the first transmitted symbol, i.e., $b_1(1)$, or a backward SIC started from the last transmitted symbol, i.e., $b_K(N)$. 

For example, for forward processing, $b_1(1)$ can be decoded by using $y_1(1)$ without interference, then $b_2(1)$ can be decoded by cancelling the interference of $b_1(1)$ from $y_2(1)$, and so on. The same procedure can be performed backwards. One can also combine forward and backward operations. However, when hard decisions are used, such a combination will not result in a noticeable gain. On the other hand, by using soft decisions, combining the forward and backward operations will improve the results as explained in the following section. 

\subsection{Forward Backward Belief Propagation Detection}
In the previous section we introduced an SIC method which was performed by passing hard decisions of previously decoded symbols to cancel the interference. In this section, we introduce a similar detection method which passes likelihood values. By using likelihood values, instead of hard decisions, performance can be improved as shown by simulation results. Additionally, this method provides the opportunity to exploit benefits of backward processing as well. We explain the strategy of decoding for BPSK modulation and $K=2$, but it can be also generalized to other modulations and other values of $K$. We also assume that transmitted symbols have the same prior probabilities and calculate the conditional probabilities as follows:
\begin{flalign}
&\left\{
\begin{array}{ll}
a=P(y_1(1)|b_1(1)=0)=\frac{1}{\sqrt{2\pi\rho_1\sigma^2} }\exp{\left(-\frac{|y_1(1)+h_1\rho_1|^2}{2\rho_1\sigma^2}\right)}\\
\nonumber
b=P(y_1(1)|b_1(1)=1)=\frac{1}{\sqrt{2\pi\rho_1\sigma^2} }\exp{\left(-\frac{|y_1(1)-h_1\rho_1|^2}{2\rho_1\sigma^2}\right)}\\
\nonumber
P_0^{fw}(b_1(1))=P(b_1(1)=0|y_1(1))=\frac{a}{a+b}\\
\nonumber
P_1^{fw}(b_1(1))=P(b_1(1)=1|y_1(1))=\frac{b}{a+b}
\end{array}\right.\\
&\left\{
\begin{array}{ll}
c=P(y_2(1)|b_2(1)=0,y_1(1))=\\
P_0^{fw}(b_1(1))\frac{1}{\sqrt{2\pi\rho_2\sigma^2}}\exp{\left(-\frac{|y_2(1)+h_1\rho_1+h_2\rho_2|^2}{2\rho_2\sigma^2}\right)}+\\
P_1^{fw}(b_1(1))\frac{1}{\sqrt{2\pi\rho_2\sigma^2}}\exp{\left(-\frac{|y_2(1)-h_1\rho_1+h_2\rho_2|^2}{2\rho_2\sigma^2}\right)}\\
\nonumber
d=P(y_2(1)|b_2(1)=1,y_1(1))=\\
P_0^{fw}(b_1(1))\frac{1}{\sqrt{2\pi\rho_2\sigma^2}}\exp{\left(-\frac{|y_2(1)+h_1\rho_1-h_2\rho_2|^2}{2\rho_2\sigma^2}\right)}+\\
P_1^{fw}(b_1(1))\frac{1}{\sqrt{2\pi\rho_2\sigma^2}}\exp{\left(-\frac{|y_2(1)-h_1\rho_1-h_2\rho_2|^2}{2\rho_2\sigma^2}\right)}\\
\nonumber
P_0^{fw}(b_2(1))=P(b_2(1)=0|y_1(1),y_2(1))=\frac{c}{c+d}\\
\nonumber
P_1^{fw}(b_2(1))=P(b_2(1)=1|y_1(1),y_2(1))=\frac{d}{c+d}
\end{array}\right.
\end{flalign}
where $\rho_i=\boldsymbol{U_{11}}(i,i)$. Using these successive calculations, $P_0^{fw}(b_k(n))$ and $P_1^{fw}(b_k(n))$ can be found for all values of $1\leq n\leq N$ and $1\leq k\leq K$. As explained before, due to the structure of the sampling method in Fig. \ref{asynch}, the last transmitted symbol can also be detected without interference and the same procedure can be applied backward to find $P_0^{bw}(b_k(n))$ and $P_1^{bw}(b_k(n))$. Using either of these likelihood sets as a detection metric will result in an improvement over the hard-decision SIC method that was presented in the previous section. Moreover, the performance can even surpass the performance of the synchronous ML detection if we use forward and backward operations together and define the detection metric as:
\begin{flalign}
\nonumber
P_0(b_k(n))=P_0^{fw}(b_k(n))P_0^{bw}(b_k(n))\\
\nonumber
P_1(b_k(n))=P_1^{fw}(b_k(n))P_1^{bw}(b_k(n))
\end{flalign}
Simulation results are presented in Section \ref{section:Simulation Results}.

\subsection{Zero Forcing}
One of the well-known linear multiuser receivers is the ZF receiver which cancels the interference caused by the other users in the expense of enhancing the noise. In a synchronized system, we need at least K receive antennas to be able to perform ZF detection; however, by exploiting asynchrony, we can perform ZF with only one receive antenna \cite{shao2007performance}, \cite{das2011mimo}. To have a fair comparison with the synchronous case, we study the system model when multiple receive antennas are used at the receiver. By stacking output samples of all receive antennas together we can represent the system model as follows:

\begin{flalign}
\nonumber
&\begin{bmatrix}
\boldsymbol{y_1}\\ 
\boldsymbol{y_2}\\ 
\vdots\\ 
\boldsymbol{y_M}
\end{bmatrix}=
\begin{bmatrix}
\boldsymbol{U}&\boldsymbol{0} & \dots &\boldsymbol{0}\\ 
\boldsymbol{0}& \boldsymbol{U}& \dots &\boldsymbol{0} \\ 
\vdots& \vdots & \ddots&\vdots \\ 
\boldsymbol{0}& \dots & \boldsymbol{0}&\boldsymbol{U} 
\end{bmatrix}\begin{bmatrix}
\boldsymbol{H_1}\\ 
\boldsymbol{H_2}\\ 
\vdots\\ 
\boldsymbol{H_M}
\end{bmatrix}\boldsymbol{b}+\begin{bmatrix}
\boldsymbol{v_1}\\ 
\boldsymbol{v_2}\\ 
\vdots\\ 
\boldsymbol{v_M}
\end{bmatrix}\\
\nonumber
&\boldsymbol{y_{tot}}=\boldsymbol{U_{tot}}\boldsymbol{H_{tot}}\boldsymbol{b}+\boldsymbol{v_{tot}}\\
&\boldsymbol{y_{tot}}=\boldsymbol{L_{tot}b}+\boldsymbol{v_{tot}}
\nonumber
\end{flalign}
where $M$ is number of receive antennas. Then, the zero-forcing detector is defined as:
\begin{flalign}\label{eq:ZF}
\boldsymbol{\tilde{y}}=(\boldsymbol{L_{tot}}^H\boldsymbol{\Sigma_{tot}}^{-1}\boldsymbol{L_{tot}})^{-1}\boldsymbol{L_{tot}}^H\boldsymbol{\Sigma_{tot}}^{-1}\boldsymbol{y_{tot}}=\boldsymbol{b}+\boldsymbol{\tilde{v}}
\end{flalign}
where $\boldsymbol{\Sigma_{tot}}=\boldsymbol{I_M} \otimes \boldsymbol{\Sigma}$, and $\boldsymbol{\Sigma}$ is an $NK\times NK$ diagonal matrix representing the covariance matrix of noise samples $\boldsymbol{v_i}$, $1\leq i \leq M$. The noise enhancement factor is $(\boldsymbol{L_{tot}}^H\boldsymbol{\Sigma_{tot}}^{-1}\boldsymbol{L_{tot}})^{-1}$, which affects the receiver performance and will be studied in the following section. 

\section{Performance Analysis}

All asynchronous receivers presented in the previous section provide full diversity. Because the ZF receiver has the worst performance among all introduced receivers, we only need to show full diversity for the ZF receiver. The system represented in Eq. (\ref{eq:ZF}) consists of $N K$ subchannels, each of them having SNR of $\frac{E[|b_k(i)|^2]}{\boldsymbol{COV_{\tilde{v}}}(i,i)}$, \ $1\leq i\leq NK$, where $\boldsymbol{COV_{\tilde{v}}}$ can be calculated as:
\begin{flalign}
\boldsymbol{COV_{\tilde{v}}}=E[\boldsymbol{\tilde{v}}\boldsymbol{\tilde{v}}^H]&=\sigma^2(\boldsymbol{L_{tot}}^H\boldsymbol{\Sigma_{tot}}^{-1}\boldsymbol{L_{tot}})^{-1}\\
\label{eq:manip}
&=\sigma^2(\sum_{i=1}^{M}{\boldsymbol{L_i}^H\boldsymbol{\Sigma}^{-1}\boldsymbol{L_i}})^{-1}\\
\label{eq:id}
&=\sigma^2(\sum_{i=1}^{M}{\boldsymbol{H_i}^*\boldsymbol{R}\boldsymbol{H_i}})^{-1}
\end{flalign}
where $\boldsymbol{L_i}=\boldsymbol{UH_i}$. In the derivation of $\boldsymbol{COV_{\tilde{v}}}$, Eq. (\ref{eq:manip}) is found by some matrix manipulation and Eq. (\ref{eq:id}) is obtained by using the fact that $ \boldsymbol{U}^H\boldsymbol{\Sigma}^{-1}\boldsymbol{U}=\boldsymbol{R}$. This identity can be simply verified by examining matrices defined in Eqs. (\ref{eq:sampling1_rect}) and (\ref{eq:sampling2_2}). Unfortunately, due to the complex structure of $\left(\sum_{i=1}^{M}{\boldsymbol{H_i}^*\boldsymbol{R}\boldsymbol{H_i}}\right)^{-1}$ for $M \geq 1$, finding the exact expression of bit error rate (BER) for $M\geq 1$ is not easy. We derive an upper bound on BER by finding an upper bound on the diagonal elements of $\boldsymbol{COV_{\tilde{v}}}$ and show that full diversity is achieved. 
Because $\boldsymbol{R}$ is positive definite, for every $1\leq i \leq M$, $\boldsymbol{H_i^*RH_i}$ is also positive definite. Therefore, we can apply the following lemma.

\begin{lemma}
For n positive definite matrices $\boldsymbol{A_i}, 1\leq i \leq n $, we have:
\begin{flalign}
(\sum_{i=1}^{n}{\boldsymbol{A_i}})^{-1}\leq \sum_{i=1}^{n}{\boldsymbol{A_i}^{-1}}
\end{flalign}
where $\boldsymbol{B}\leq \boldsymbol{C}$ means that $\boldsymbol{C}-\boldsymbol{B}$ is positive semidefinite. 
\end{lemma}
\begin{IEEEproof}
This lemma is a straightforward result of the following inequality, which can be found in \cite{horn2012matrix}.
\begin{flalign}
\nonumber
(\boldsymbol{A}+\boldsymbol{B})^{-1}\leq \boldsymbol{A}^{-1}\ \ \ \boldsymbol{A},\boldsymbol{B}: \text{ positive definite matrices }
\end{flalign}
\end{IEEEproof}

As a result, we can conclude that $\boldsymbol{COV_{\tilde v}} \leq \sigma^2\sum_{i=1}^{M}{\left(\boldsymbol{H_i}^*\boldsymbol{ R} \boldsymbol{H_i}\right)^{-1}}$. This inequality implies that the diagonal elements of the covariance matrix of noise are upper bounded as follows:
\begin{flalign}\label{eq:ineq}
\boldsymbol{COV_{\tilde v}}(i,i) \leq \frac{\sigma^2\boldsymbol{R}^{-1}(i,i)}{\sum_{j=1}^{M}{|h_{\left(1+(i-1) mod K\right),j}|^2}}\ \ \ 1\leq i \leq NK
\end{flalign}
where $h_{k,m}$ is the channel coefficient between User $k$ and Receive Antenna $m$.\\
The BER expression for an AWGN channel with average transmit power of ${E[|b_k(i)|^2]}$ and noise variance of $\frac{\sigma^2\boldsymbol{R}^{-1}(i,i)}{\sum_{j=1}^{M}{|h_{((1+(i-1) mod K),j)}|^2}}$ is equal to:
\begin{align}\label{eq:BER2}
p_i&=\frac{\sqrt{\frac{\delta_0 2}{\pi\boldsymbol{R}^{-1}(i,i)}}}{2\left(1+\frac{\delta_0 2}{\boldsymbol{R}^{-1}(i,i)}\right)^{M+\frac{1}{2}}}\times \\
\nonumber
&\frac{\Gamma(M+\frac{1}{2})}{\Gamma(M+1)}\times \leftidx{_2}{F}{_1}(1,M+\frac{1}{2};M+1;\frac{1}{1+\frac{\delta_0 2}{ \boldsymbol{R}^{-1}(i,i)}})
\end{align}
where $\delta_0=\frac{E[|b_k(i)|^2]}{\sigma^2}$. The details of derivation can be found in Appendix \ref{section:BER}. Due to having the same average transmit power and a lower noise variance, we conclude that
BER for each subchannel  is upper bounded by $p_i$, i.e., $P_i\leq p_i$.
If we define $D_i=-\lim_{\delta_0 \rightarrow \infty}{\frac{\log{P_i}}{\log{\delta_0}}}$ and $d_i=-\lim_{\delta_0 \rightarrow \infty}{\frac{\log{p_i}}{\log{\delta_0}}}$, it is clear that $D_i \geq d_i$. By using the fact that the hypergeometric function of form $\leftidx{_2}{F}{_1}(1,m+\frac{1}{2};m+1;\frac{1}{1+c})$ converges to one as $c$ grows large \cite{integral2}, we can calculate that $d_i=M$. Therefore, the diversity of the $i$th subchannel  is greater than or equal to $M$. On the other hand, $M$ is the maximum available diversity for this system, which completes the proof of achieving full diversity, i.e. $D_i=M$.

\subsection{Analyzing Asymptotic Performance for Large Number of Receive Antennas}

In this section, we show that when $M$ goes to infinity, the correlation between noise samples and the effect of fading coefficients vanishes. For normalizing purposes, instead of using Eq. (\ref{eq:ZF}), we perform ZF by multiplying output samples by $\sqrt M(\boldsymbol{L_{tot}}^H\boldsymbol{\Sigma_{tot}}^{-1}\boldsymbol{L_{tot}})^{-1}\boldsymbol{L_{tot}}^H\boldsymbol{\Sigma_{tot}}^{-1}$in this section. Then, the inverse of noise covariance will be equal to $\frac{1}{\sigma^2M}\sum_{i=1}^{M}{\boldsymbol{H_i}^*\boldsymbol{R}\boldsymbol{H_i}}$, which can be represented as:
\begin{flalign}
\boldsymbol{COV_{\tilde{v}}}^{-1}=\frac{1}{\sigma^2}\boldsymbol{R}\circ (\boldsymbol{J_N} \otimes \boldsymbol{\tilde H})
\end{flalign}
where $(\circ)$ is Hadamard product and $\boldsymbol{J_N}$ is the $N\times N$ all-ones matrix. $\boldsymbol{\tilde{H}}$ is also  defined as follows:
\begin{flalign}
\nonumber
\frac{1}{M}\left(\begin{smallmatrix}
\sum_{m=1}^{M}{|h_{1,m}|^2}& \sum_{m=1}^{M}{h_{1,m}h^*_{2,m}}& \dots & \sum_{m=1}^{M}{h_{1,m}h^*_{K,m}}\\ 
\sum_{m=1}^{M}{h_{2,m}h^*_{1,m}} &\sum_{m=1}^{M}{|h_{1,m}|^2}& \dots & \sum_{m=1}^{M}{h_{2,m}h^*_{K,m}}\\ 
\vdots & \ddots&\ddots & \vdots\\
\sum_{m=1}^{M}{h_{K,m}h^*_{1,m}} &\sum_{m=1}^{M}{h_{K,m}h^*_{2,m}} &\dots &\sum_{m=1}^{M}{|h_{K,m}|^2}
\end{smallmatrix}\right)
\end{flalign}
Then, by using the law of large numbers, it can be shown that $\boldsymbol{\tilde H} \rightarrow \boldsymbol{I_K}$ as $M \rightarrow \infty$ \cite{cramer2004random}. The immediate result is that $\boldsymbol{COV_{\tilde{v}}}^{-1}$ and $\boldsymbol{COV_{\tilde{v}}}$ approaches to $\frac{1}{\sigma^2}\boldsymbol{I_{NK}}$ and $\sigma^2\boldsymbol{I_{NK}}$, respectively. This result implies that at the output of the ZF receiver, noise samples are independent and SNR for each subchannel is a fixed value of $M\delta_0$, independent of channel coefficients. In other words, the effect of fading coefficients and correlation between noise samples will vanish.

\subsection{Effect of Time Delays on Performance}\label{subsection:delay}
In this section, we calculate the optimal values of delays for the ZF detection in order to achieve the lowest average BER with one receive antenna at high SNR. Because for $M=1$ the inequality in Eq. (\ref{eq:ineq}) turns into equality, the exact BER expression for each subchannel can be obtained as:
\begin{align}
\nonumber
P_i&=\\
\nonumber
&\frac{\sqrt{\frac{\delta_0 2}{\pi \boldsymbol{R}^{-1}(i,i)}}}{2\left(1+\frac{\delta_0 2}{\boldsymbol{R}^{-1}(i,i)}\right)^{3/2}}\frac{\Gamma(3/2)}{\Gamma(2)}\  \leftidx{_2}{F}{_1}(1,3/2;2;\frac{1}{1+\frac{\delta_0 2}{ \boldsymbol{R}^{-1}(i,i)}})
\end{align}
Approximating $P_{avg}$ at high SNR for one receive antenna results in: (see Appendix \ref{section:average} for more details)
\begin{align}
\nonumber
\widetilde{P_{avg}}&=\frac{1}{4\sqrt{\pi}NK}\frac{\Gamma(3/2)}{\Gamma(2)} \times \frac{\sum_{i}\boldsymbol{R}^{-1}(i,i)}{\delta_0}, 
\end{align}
For a fixed number of users and frame length, in order to maximize $\widetilde{P_{avg}}$, we need to maximize the trace($\boldsymbol{R}^{-1}$) which is related to time delays between different users. In what follows, we derive the relationship between the trace($\boldsymbol{R}^{-1}$) and time delays, and consequently find optimum time delays. 
\begin{lemma} \label{lemma:trace}
the sum of the diagonal elements of the inverse of matrix $\boldsymbol{R}$ is equal to: 
\begin{flalign}\label{eq:sum}
\nonumber
\text{trace}(\boldsymbol{R}^{-1})=&\frac{(N-1)(N+1)}{3(1+\tau_1-\tau_K)}+\frac{2N+1}{3(N+1+\tau_1-\tau_K)}\\
&+\frac{N(N+2)}{3} \sum^{K-1}_{i=1}{\frac{1}{\tau_{i+1}-\tau_i }}
\end{flalign}
\end{lemma}
The proof is presented in Appendix \ref{section:trace}.
\begin{theorem} \label{theorem:optimum}
The optimum time delays which result in the lowest average BER for ZF detection at high SNR are: ($\tau_1$ is assumed to be zero)
\begin{flalign}\label{eq:recursive}
\tau_{i-1}=\frac{i-2}{i-1} \times \tau_{i} \ \ \ 3 \leq i \leq K
\end{flalign}
Also $\tau_K$ is found by solving the following equation:
\begin{flalign}\label{eq:best}
&A\tau_K^4+B\tau_K^3+C\tau_K^2+D\tau_K+E=0\\
\nonumber
&\text{where}\\
\nonumber
&A=(1-(K-1)^2)\frac{(N+2)}{3}.\\
\nonumber
&B=\frac{-2}{3}(1-(K-1)^2)N^2+2(4(K-1)^2-1)\frac{(N+1)}{3}.\\
\nonumber
&C=\frac{1}{3}(1-(K-1)^2)N^3+\frac{2}{3}(1-4(K-1)^2)N^2-\\
\nonumber
& \ \ \ \ \ \ 2(K-1)^2(3N+2).\\
\nonumber
&D=\frac{2}{3}(K-1)^2(N^3+5N^2+8N+4).\\
\nonumber
&E=-\frac{1}{3}(K-1)^2(N^3+4N^2+5N+2).
\end{flalign}
\end{theorem}
The proof is easily obtained by taking the derivation of Eq. (\ref{eq:sum}) with respect to time delays. \\
For $K=2$, $A$ will be zero and Eq. (\ref{eq:best}) is a polynomial of degree 3 which has a closed-form solution as follows:
\begin{flalign}
\tau_{opt}=\frac{N+2-\sqrt[3]{N^3+1.5N^2-1.5N-1}}{3}
\end{flalign}
 where $N$ is the block length. However, for other values of $K$, Eq. (\ref{eq:best}) should be solved numerically. After finding $\tau_K$, the remaining time delays are calculated recursively using Eq. (\ref{eq:recursive}). The optimum delay values for different $K$ and $N$ values are reported in Tables \ref{table:nonlin1} and \ref{table:nonlin2}.
\begin{table}[h!]
\caption{Optimum Time Delays when $K=2$} 
\centering 
{\tabulinesep=1mm
   \begin{tabu}{c c c c c c}
\hline\hline 
Case & N=10 & N=32 & N=64 & N=128 & N $\rightarrow\infty$ \\ 
\hline 
K=2 & 0.5240 & 0.5077 &  0.5039 & 0.5019 &0.5 \\ 
\hline 
\end{tabu}}
\label{table:nonlin1} 
\end{table}
\begin{table}[h!] \label{table:II}
\caption{Optimum delays when $N=128$} 
\centering 
{\tabulinesep=1mm
   \begin{tabu}{c c }
\hline\hline 
Case & N=128 \\ 
\hline 
K=4 & [0.2505,0.5010,0.7514] \\ 
\hline 
K=6 & [0.1669,0.3338,0.5006,0.6675,0.8344] \\ 
\hline 
K=8 &[0.1251,0.2502,0.3754,0.5004,0.6256,0.7507,0.8758] \\ 
\hline 
\end{tabu}}
\label{table:nonlin2} 
\end{table}
Optimum time delays approach uniform time delays, i.e, $\tau_{k}=\frac{k-1}{K}$, $2 \leq k \leq K$, as $N$ increases. The effects of time delay values on the performance are studied numerically in the following section. 
\section{Simulation Results}\label{section:Simulation Results}
In this section, we provide simulation results in order to validate our theoretical results and compare different methods. In all simulations, channel coefficients are independent Rayliegh fadings with variance one, fixed during the block and changing independently for each block. All users have the same average power of one and variance of noise ($\sigma^2$) is equal to $10^\frac{-\text{SNR}}{10}$ where SNR is in dB. To avoid inter-block interference, the last symbol of each block should be idle for asynchronous methods. This will reduce spectral efficiency, but it is negligible for large block lengths. In all simulations, the block length is 128 and the time delays are uniform except in the case where we report the time delays to study their effects on the performance. The number of users and the number of receive antennas is denoted by $K$ and $M$, respectively. When $M$ is not specified, the assumption is that only one receive antenna is used. Transmitted symbols are chosen from BPSK modulation and the comparing criterion is the average bit error rate among all the users.\\

\begin{figure}[h!]
\centering
\includegraphics[width=3in]{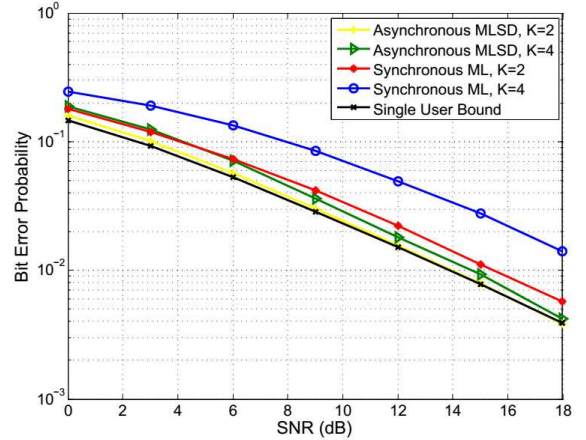}
\caption{Comparing asynchronous MLSD and synchronous ML}\label{fig:MLSD}
\end{figure}

\begin{figure}[h!]
\centering
\includegraphics[width=3in]{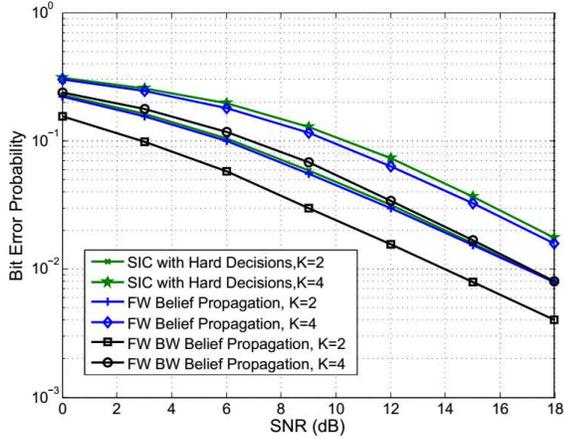}
\caption{Performance of SIC method with hard decisions and soft decisions}\label{fig:BP}
\end{figure}

In Fig. \ref{fig:MLSD},we compare the performance of the asynchronous MLSD method with that of the synchronous ML. Asynchronous MLSD outperforms synchronous ML detection with similar complexity. Fig. \ref{fig:MLSD} also includes the single-user bound for a better comparison. As can be seen in the figure, asynchronous MLSD for $K=2$ achieves performance of the single user system at high SNR.\\
Fig. \ref{fig:BP} shows the performance of different SIC methods presented in Section \ref{section:SIC}. Our new forward backward belief propagation method using the sampling method in Fig. \ref{asynch} improves the performance of traditional SIC method by about 3 dB. \\
\begin{figure}[h!]
\centering
\includegraphics[width=3in]{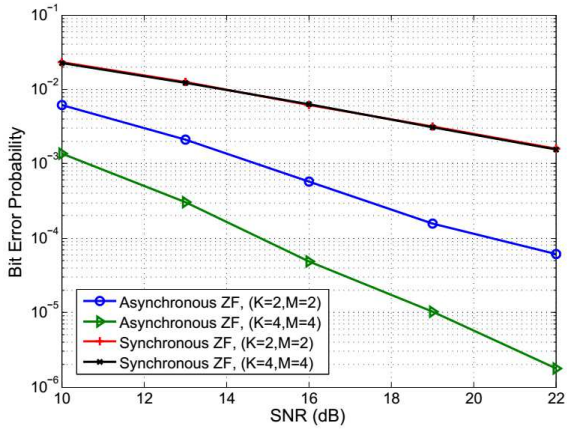}
\caption{Comparing synchronous and asynchronous ZF}\label{fig:ZF}
\end{figure}
Fig. \ref{fig:ZF} compares the performance of the synchronous and asynchronous ZF detectors. Although asynchronous ZF is possible with one receive antenna, for fair comparison, we consider the cases where the number of receive antennas and users are the same. Since all users are assumed to have the same transmit power, synchronous ZF for $(K=2,M=2)$ and $(K=4,M=4)$ provides the same performance and both of them have diversity of one. However, for asynchronous ZF detection, diversity of $2$ and $4$ is achieved for $(K=2,M=2)$ and $(K=4,M=4)$, respectively. This is due to the sampling diversity as discussed earlier.

\begin{figure}[h!]
\centering
\includegraphics[width=3in]{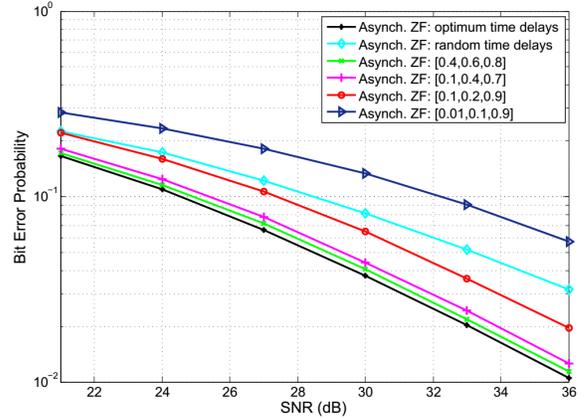}
\caption{Effect of time delays in asynchronous ZF detection for $K=4$}\label{fig:ZF_delay}
\end{figure}

\begin{figure}[h!]
\centering
\includegraphics[width=3in]{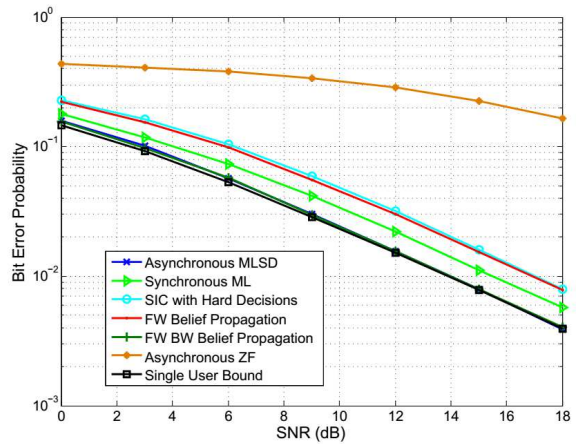}
\caption{Comparison of all detection methods for $K=2$}\label{fig:all}
\end{figure}

We study the effects of time delay values on the performance of a ZF system with $K=4$ users and one receive antenna in Fig. \ref{fig:ZF_delay}. Note that a synchronous ZF solution does not exist in this case as we need at least $M=4$ receive antennas. We show the results for six different sets of time delays. For optimum time delays we use the result of Section \ref{subsection:delay} as reported in Table \ref{table:nonlin2}. The curve associated with random time delays represents the average performance over uniformly distributed random time delays. The remaining sets of time delays are specified in the figure. The optimum time delays and time delays of $[0.01,0.1,0.9]$ have the best and worst performances, respectively. They also have the lowest and the highest trace($\boldsymbol{R}^{-1}$), respectively, which are presented along with other sets of time delays in  Table \ref{table:III}. As can be seen, a lower trace($\boldsymbol{R}^{-1}$) results in a better performance. This observation is in line with the analysis in Section \ref{subsection:delay} where trace($\boldsymbol{R}^{-1}$) was introduced as a criterion to compare the performance of different time delays.\\
\begin{table}[h!] 
\caption{Comparing trace($\boldsymbol{R}^{-1}$) for different time delays in Fig. \ref{fig:ZF_delay}} 
\centering 
{\tabulinesep=1.2mm
   \begin{tabu} { |c |c |}
\hline 
Time delays &  trace($\boldsymbol{R}^{-1}$) \\ 
\hline 
$[0.2505,0.5010,0.7514]$ & $8.8404\times 10^4$ \\ 
\hline 
$[0.4,0.6,0.8]$ & $9.6639 \times 10^4$\\ 
\hline 
$[0.1,0.4,0.7]$ & $1.1065\times 10^5$\\
\hline 
 $[0.1,0.2,0.9]$ & $1.7347 \times 10^5$\\ 
\hline 
$[0.01,0.1,0.9]$ & $6.7784 \times 10^5$\\ 
\hline 
\end{tabu}}
\label{table:III}
\end{table}
Finally, to compare different methods with each other, we include the performance of all detection methods for $K=2$ in Fig. \ref{fig:all}. Both MLSD and forward-backward BP detection methods not only outperform the synchronous ML detection, but also achieve the performance of the single user system. In addition, the low complexity method of SIC with hard decisions also provides good performance.
\section{Conclusion}
In this paper, we studied benefits of asynchrony when multiple users are sending data simultaneously to a common receiver. Instead of treating asynchrony as a disruptive factor, we exploited it as an additional resource to cancel interference. We have shown that asynchrony between data streams adds a favorable ISI which makes interference cancellation possible. It also introduces memory to the system which can be exploited by methods like maximum-likelihood sequence detection. In addition to MLSD, a novel forward-backward belief propagation detection method was presented and this method outperforms synchronous ML detection. Exact BER expression for ZF detection was derived and it was verified that a diversity equal to the number of receive antennas is achievable by asynchronous transmission.
\vspace*{-0.5em}
\bibliographystyle{IEEEtran}
\bibliography{reference}

\begin{thebibliography}{10}
\providecommand{\url}[1]{#1}
\csname url@samestyle\endcsname
\providecommand{\newblock}{\relax}
\providecommand{\bibinfo}[2]{#2}
\providecommand{\BIBentrySTDinterwordspacing}{\spaceskip=0pt\relax}
\providecommand{\BIBentryALTinterwordstretchfactor}{4}
\providecommand{\BIBentryALTinterwordspacing}{\spaceskip=\fontdimen2\font plus
\BIBentryALTinterwordstretchfactor\fontdimen3\font minus
  \fontdimen4\font\relax}
\providecommand{\BIBforeignlanguage}[2]{{%
\expandafter\ifx\csname l@#1\endcsname\relax
\typeout{** WARNING: IEEEtran.bst: No hyphenation pattern has been}%
\typeout{** loaded for the language `#1'. Using the pattern for}%
\typeout{** the default language instead.}%
\else
\language=\csname l@#1\endcsname
\fi
#2}}
\providecommand{\BIBdecl}{\relax}
\BIBdecl

\bibitem{honig2009advances}
M.~L. Honig, \emph{{Advances in multiuser detection}}.\hskip 1em plus 0.5em
  minus 0.4em\relax Wiley Online Library, 2009.

\bibitem{verdu1998multiuser}
S.~Verdu, \emph{{Multiuser detection}}.\hskip 1em plus 0.5em minus 0.4em\relax
  Cambridge University Press, 1998.

\bibitem{moshavi1996multi}
S.~Moshavi, ``{Multi-user detection for DS-CDMA communications},'' \emph{IEEE
  Communications Magazine,}, vol.~34, no.~10, pp. 124--136, 1996.

\bibitem{naguib1998applications}
A.~F. Naguib, N.~Seshadri, and A.~R. Calderbank, ``{Applications of space-time
  block codes and interference suppression for high capacity and high data rate
  wireless systems},'' \emph{Record of the 32nd Asilomar Conference on Signals,
  Systems \& Computers.}, vol.~2, pp. 1803--1810, 1998.

\bibitem{kazemitabar2008multiuser}
J.~Kazemitabar and H.~Jafarkhani, ``{Multiuser interference cancellation and
  detection for users with more than two transmit antennas},'' \emph{IEEE
  Transactions on Communications}, vol.~56, no.~4, pp. 574--583, 2008.

\bibitem{shao2007performance}
S.~Shao, Y.~Tang, T.~Kong, K.~Deng, and Y.~Shen, ``Performance analysis of a
  modified v-blast system with delay offsets using zero-forcing detection,''
  \emph{IEEE Transactions on Vehicular Technology}, vol.~56, no.~6, pp.
  3827--3837, 2007.

\bibitem{das2011mimo}
A.~Das and B.~D. Rao, ``{MIMO Systems with Intentional Timing Offset},''
  \emph{EURASIP Journal on Advances in Signal Processing}, vol. 2011, no.~1,
  pp. 1--14, 2011.

\bibitem{poorkasmaei2015asynchronous}
S.~Poorkasmaei and H.~Jafarkhani, ``{Asynchronous Orthogonal Differential
  Decoding for Multiple Access Channels},'' \emph{IEEE Transactions on Wireless
  Communications}, vol.~14, no.~1, pp. 481--493, 2015.

\bibitem{avendi2015differential}
M.~Avendi and H.~Jafarkhani, ``{Differential distributed space-time coding with
  imperfect synchronization in frequency-selective channels},'' \emph{IEEE
  Transactions on Wireless Communications}, vol.~14, no.~4, pp. 1811--1822,
  2015.

\bibitem{verdu}
S.~Verdu, ``{The capacity region of the symbol-asynchronous Gaussian
  multiple-access channel},'' \emph{IEEE Transactions on Information Theory},
  vol.~35, no.~4, pp. 733--751, 1989.

\bibitem{proakis2001intersymbol}
J.~G. Proakis, \emph{Intersymbol interference in digital communication
  systems}.\hskip 1em plus 0.5em minus 0.4em\relax Wiley Online Library, 2001.

\bibitem{horn2012matrix}
R.~A. Horn and C.~R. Johnson, \emph{{Matrix analysis}}.\hskip 1em plus 0.5em
  minus 0.4em\relax Cambridge University Press, 2012.

\bibitem{integral2}
R.~Xu and F.~Lau, ``{Performance analysis for MIMO systems using zero forcing
  detector over fading channels},'' \emph{IEEE Proceedings-Communications},
  vol. 153, no.~1, pp. 74--80, 2006.

\bibitem{cramer2004random}
H.~Cram{\'e}r, \emph{{Random variables and probability distributions}}.\hskip
  1em plus 0.5em minus 0.4em\relax Cambridge University Press, 2004.

\bibitem{integral1}
T.~Eng and L.~B. Milstein, ``{Coherent DS-CDMA performance in Nakagami
  multipath fading},'' \emph{IEEE Transactions on Communications}, vol.~43, no.
  2/3/4, pp. 1134--1143, 1995.

\bibitem{dow2008explicit}
M.~Dow, ``{Explicit inverses of Toeplitz and associated matrices},''
  \emph{ANZIAM Journal}, vol.~44, pp. 185--215, 2008.

\bibitem{bernstein2009matrix}
D.~S. Bernstein, \emph{{Matrix mathematics: theory, facts, and
  formulas}}.\hskip 1em plus 0.5em minus 0.4em\relax Princeton University
  Press, 2009.

\end{thebibliography}

\appendices 
\section{Derivation of Bit Error Rate (BER) Expression}\label{section:BER}
For an AWGN channel with an average transmit power of ${E[|b_k(i)|^2]}$ and noise variance of $\frac{\sigma^2\boldsymbol{R}^{-1}(i,i)}{\sum_{j=1}^{M}{|h_{(1+(i-1) mod K),j}|^2}}$, the post SNR at the receiver can be expressed as:
\begin{flalign}
\delta_i =\frac{\delta_0 \sum_{j=1}^{M}{|h_{(1+(i-1) mod K),j}|^2}}{\boldsymbol{R}^{-1}(i,i)}
\end{flalign}
where $\delta_0 =\frac{E[|b_k(i)|^2]}{\sigma^2}$. We know that $|h_{i,j}|^2$ follows a chi-squared distribution with two degrees of freedom for all $i$s and $j$s. Therefore, $\sum_{j=1}^{M}{|h_{(1+(i-1) mod K),j}|^2}$ is chi-squared distributed with $2M$ degrees of freedom. As a result, the distribution of $\delta_i$ can be calculated as follows:
\begin{flalign}
P_{\delta_i}(\delta)=\frac{\boldsymbol{R}^{-1}(i,i)}{\delta_0 } \frac{\left(\frac{\boldsymbol{R}^{-1}(i,i)}{\delta_0}\delta\right)^{M-1} \exp{\left(-\frac{\boldsymbol{R}^{-1}(i,i)}{\delta_0 2}\delta\right)}}{2^{M}\Gamma(M)} 
\end{flalign}
where $\Gamma(.)$ is the Gamma function. 
For a specific value of SNR, BER varies according to the modulation. We assume that BPSK is used, however, extension to other modulations is straightforward. Based on this assumption, the BER for a given value of SNR, e.g., $\delta$ is equal to $Q(\sqrt{2\delta})$. The next step is to calculate the following integral:

\begin{flalign}
\nonumber
p_i&=\int^{\infty }_{0}{Q(\sqrt{2\delta})P_{\delta_i}(\delta)d\delta}
\end{flalign}
The integral of $\frac{a^m}{\Gamma(m)}\int_{0}^{\infty}{\exp{(-az)}z^{m-1}Q(\sqrt{bz})dz}$ has a closed-form of: 
\begin{align}
\nonumber
\frac{\sqrt{b/2\pi a}}{2\left(1+\frac{b}{2a}\right)^{m+1/2}}\frac{\Gamma(m+1/2)}{\Gamma(m+1)} \leftidx{_2}{F}{_1}(1,m+\frac{1}{2};m+1;\frac{1}{1+\frac{b}{2a}})
\end{align}
where $\leftidx{_2}{F}{_1}(q,w;e;r)$ is the hypergeometric function \cite{integral1}. Therefore, the bit error rate, i.e., $p_i$ is equal to:
\begin{align}
\label{eq:BER}
p_i&=\frac{\sqrt{\frac{\delta_0 2}{\pi\boldsymbol{R}^{-1}(i,i)}}}{2\left(1+\frac{\delta_0 2}{\boldsymbol{R}^{-1}(i,i)}\right)^{M+\frac{1}{2}}}\times \\
\nonumber
&\frac{ \Gamma(M+\frac{1}{2})}{\Gamma(M+1)}\times \leftidx{_2}{F}{_1}(1,M+\frac{1}{2};M+1;\frac{1}{1+\frac{\delta_0 2}{ \boldsymbol{R}^{-1}(i,i)}})
\end{align}

\section{Average BER and Its Approximation at High SNR}\label{section:average}
In Eq. (\ref{eq:BER}), $p_i$ depends on $\boldsymbol{R}^{-1}(i,i)$ which varies for different values of $i$, and therefore each subchannel has a different BER. This is unlike the synchronous ZF, where all resulting subchannels have the same performance. In order to evaluate the performance of the entire system, we define the average BER performance as follows:
\begin{flalign}
 p_{avg}=\frac{\sum_{i=1}^{NK}{p_i}}{NK} 
\end{flalign}
Since $p_{avg}$ is not tractable, we approximate it at high SNR, using the fact that $\leftidx{_2}{F}{_1}(1,m+\frac{1}{2};m+1;\frac{1}{1+c})$ converges to one as $c$ grows large \cite{integral2}. Hence, at high SNR, $p_{avg}$ can be approximated as follows:
\begin{flalign}
\label{eq:approx}
\widetilde{p_{avg}}&=\text{Const} \ \times \frac{\sum_{i=1}^{NK}(\boldsymbol{R}^{-1}(i,i))^M}{\delta^M_0}
\end{flalign}
where the constant value is equal to $\frac{1}{2^{\left(M+1\right)}NK\sqrt{\pi}} \frac{\Gamma(M+\frac{1}{2})}{\Gamma(M+1)} $.
\section{Proof of Lemma \ref{lemma:trace}}\label{section:trace}
When the frame length is $N$, we denote $\boldsymbol{R}$ by $\boldsymbol{R^N}$. Then, we prove by induction that, for all $N\in \mathbb{Z}_+$,
\begin{flalign}\label{eq:induction}
\nonumber
\text{trace}((\boldsymbol{R^N})^{-1})=&\frac{(N-1)(N+1)}{3(1+\tau_1-\tau_K)}+\frac{2N+1}{3(N+1+\tau_1-\tau_K)}\\
&+\frac{N(N+2)}{3} \sum^{K-1}_{i=1}{\frac{1}{\tau_{i+1}-\tau_i }}
\end{flalign}

\textbf{Base case}: When $N=1$, $\boldsymbol{R^1}$ is equal to $\boldsymbol{R_{11}}$ which can be written as a generalized Fiedler’s matrix whose inverse is given by \cite{dow2008explicit}:
\begin{flalign}
\nonumber
&\boldsymbol{R_{11}}^{-1}=\\
&-\frac{1}{2}\left(\begin{smallmatrix}
d_1& \frac{1}{\tau_{2}-\tau_{1}}&\dots &0&f\\ 
\frac{1}{\tau_{2}-\tau_{1}}& d_2&\frac{1}{\tau_{3}-\tau_{2}}&\dots& 0\\ 
&\ddots&\ddots&\ddots&\\ 
0&\dots&\frac{1}{\tau_{K-1} - \tau_{K-2}}& d_{K-1}&\frac{1}{\tau_{K} - \tau_{K-1}}\\ 
f&0& \dots &\frac{1}{\tau_{K} - \tau_{K-1}}&d_{K}
\end{smallmatrix}\right)
\end{flalign}
where $f$ and $d_i$s are defined as:
\begin{flalign}
&f=\frac{1}{\tau_K-\tau_1-2}\\
&d_1=\frac{1}{\tau_1-\tau_2}-\frac{1}{\tau_1-\tau_K+2}\\
&d_K=\frac{1}{\tau_7-\tau_8}-\frac{1}{\tau_1-\tau_K+2}\\
&d_i=\frac{1}{\tau_{i-1}-\tau_i}+\frac{1}{\tau_i-\tau_{i+1}} \ \ \ \ 2 \leq i \leq K-1
\end{flalign}
Then, trace($\boldsymbol{R^{-1}_{11}}$) is equal to $\left(-\frac{1}{2}\sum^{K}_{i=1}{d_i}\right)$, which can be calculated using the above equations:
\begin{flalign}
\text{trace}(\boldsymbol{R^{-1}_{11}})=\frac{1}{(2+\tau_1-\tau_K)}+\sum^{K-1}_{i=1}{\frac{1}{\tau_{i+1}-\tau_i }}
\end{flalign}
Therefore, Eq. (\ref{eq:induction}) is true for $N=1$. \\
\textbf{Induction step}: Suppose Eq. (\ref{eq:induction}) is true for $N$. We need to show that it also holds for $N+1$, i.e.,
\begin{flalign}\label{eq:step}
\nonumber
\text{trace}((\boldsymbol{R^{(N+1)}})^{-1})=&\frac{(N)(N+2)}{3(1+\tau_1-\tau_K)}+\frac{2N+3}{3(N+2+\tau_1-\tau_K)}\\
&+\frac{(N+1)(N+3)}{3} \sum^{K-1}_{i=1}{\frac{1}{\tau_{i+1}-\tau_i }}
\end{flalign}
Because matrix $\boldsymbol{R}$ follows a recursive structure, $\boldsymbol{R^{N+1}}$can be presented as follows:
\begin{flalign}\label{eq:rec}
\nonumber
\boldsymbol{R^{N+1}}&=\begin{bmatrix}
(\boldsymbol{R^N})_{NK\times NK}& (\boldsymbol{L})_{NK\times K}\\ 
(\boldsymbol{L}^T)_{K\times NK}& (\boldsymbol{R_{11}})_{K\times K}
\end{bmatrix}
\end{flalign}
where $\boldsymbol{L}^T=[\boldsymbol{0}_{K\times K},\dots,\boldsymbol{0}_{K\times K},(\boldsymbol{R_{21}})_{K\times K}]$. For calculating the inverse of $\boldsymbol{R^{N+1}}$, we use the following lemma for matrix inversion in block form.
\begin{lemma}\label{lemma:block}
Let na $(m+n) \times (m+n)$ matrix $\boldsymbol{T}$ be partitioned into a block form:
\begin{flalign}
\nonumber
\boldsymbol{T}=\begin{bmatrix}
\boldsymbol{A}& \boldsymbol{B}\\ 
\boldsymbol{C}&\boldsymbol{D} 
\end{bmatrix}
\end{flalign}
where the $m \times m$ matrix $\boldsymbol{A}$ and $n \times n$ matrix $\boldsymbol{D}$ are invertible. Then, we have:
\begin{flalign}
\nonumber
\boldsymbol{T}^{-1}&=\begin{bmatrix}
\boldsymbol{M}^{-1}& -\boldsymbol{M}^{-1}\boldsymbol{B}\boldsymbol{D}^{-1}\\ 
-\boldsymbol{D}^{-1}\boldsymbol{C}\boldsymbol{M}^{-1}& \boldsymbol{D}^{-1}+\boldsymbol{D}^{-1}\boldsymbol{C}\boldsymbol{M}^{-1}\boldsymbol{B}\boldsymbol{D}^{-1}
\end{bmatrix}
\end{flalign}
where $\boldsymbol{M}=\boldsymbol{A}-\boldsymbol{BD}^{-1}\boldsymbol{C}$ \cite{bernstein2009matrix}.
\end{lemma}
Here, $\boldsymbol{A}$, $\boldsymbol{B}$, $\boldsymbol{C}$ and $\boldsymbol{D}$ are equal to $\boldsymbol{R^N}$, $\boldsymbol{L}$, $\boldsymbol{L}^T$ and $\boldsymbol{R_{11}}$, respectively. Therefore, $\boldsymbol{M}$ is equal to:
\begin{flalign}
\boldsymbol{M}=\boldsymbol{R^{N}}-\boldsymbol{L}(\boldsymbol{R_{11}})^{-1}\boldsymbol{L}^T
\end{flalign}
Now, we need to find the inverse of $\boldsymbol{M}$. By defining $\boldsymbol{Z}$ as $(\boldsymbol{R^N})^{-1}$, the inverse of $\boldsymbol{M}$ can be presented as:
\begin{flalign}\label{eq:M}
\nonumber
&\boldsymbol{M}^{-1}=\\
&\begin{bmatrix}
\boldsymbol{I}_{K}& \dots &\boldsymbol{ 0}_{K,K}&\boldsymbol{Z_{1N}}\boldsymbol{Q}(\boldsymbol{I}_K-\boldsymbol{Z_{NN}Q})^{-1}\\ 
\boldsymbol{0}_{K\times K} & \ddots &\vdots& \vdots\\ 
\vdots& \vdots & \boldsymbol{I}_{K}&\boldsymbol{Z_{(N-1)N}Q}(\boldsymbol{I}_K-\boldsymbol{Z_{NN}Q})^{-1}\\
\boldsymbol{0}_{K \times K}&\dots &\boldsymbol{0}_{K \times K}&(\boldsymbol{I}_K-\boldsymbol{Z_{NN}Q})^{-1}
\end{bmatrix}\boldsymbol{Z} 
\end{flalign}
 where $\boldsymbol{Q}=\boldsymbol{R_{12}}\boldsymbol{R_{11}}^{-1}\boldsymbol{R_{21}}$ and $\boldsymbol{Z_{ij}}$s are $K \times K$ partitioning blocks of $\boldsymbol{Z}$. Also, $\boldsymbol{I}_{k}$  and $\boldsymbol{0}_{i\times j}$ are a $k\times k$ identity matrix and a $i \times j$ all-zero matrix, respectively.\\
To show the correctness of Eq. (\ref{eq:M}), we need to take the following steps:

\textbf{Step 1}: By some calculations, it can be shown that $\boldsymbol{L}(\boldsymbol{R_{11}})^{-1}\boldsymbol{L}^T$ is equal to $\begin{bmatrix}
\boldsymbol{0}& \boldsymbol{0}\\ 
\boldsymbol{0}& \boldsymbol{Q}
\end{bmatrix}$. As a result, we have:
\begin{flalign}
&\boldsymbol{M}=\boldsymbol{R^N}-\begin{bmatrix}
\boldsymbol{0}_{(N-1)K\times (N-1)K}& \boldsymbol{0}_{(N-1)K\times K}\\ 
\boldsymbol{0}_{K\times (N-1)K}& \boldsymbol{Q}
\end{bmatrix}
\end{flalign}

\textbf{Step 2}: If we multiply both sides by $\boldsymbol{Z}$, we will have:
\begin{flalign}\label{zm}
&\boldsymbol{ZM}=\boldsymbol{I}_{NK}-\begin{bmatrix}
\boldsymbol{0}_{K\times K}& \dots & \boldsymbol{0}_{K\times K}&\boldsymbol{Z_{1N}Q}\\ 
\boldsymbol{0}_{K\times K}& \dots &\boldsymbol{0}_{K\times K}& \boldsymbol{Z_{2N}Q}\\ 
\vdots& \vdots & \vdots&\vdots\\
\boldsymbol{0}_{K\times K}&\dots &\boldsymbol{0}_{K \times K}&\boldsymbol{Z_{NN}Q} 
\end{bmatrix}
\end{flalign}

\textbf{Step 3}: We denote the right hand side of Eq. (\ref{zm}) by $\boldsymbol{X}$, then, we can conclude that the inverse of $\boldsymbol{M}$ is equal to:
\begin{flalign}\label{eq:X}
\boldsymbol{M}^{-1}=\boldsymbol{X}^{-1}\boldsymbol{Z}
\end{flalign}

\textbf{Step 4}: $\boldsymbol{X}^{-1}$ can be calculated as follows:
\begin{flalign}
\nonumber
&\boldsymbol{X}^{-1}=\\
&\begin{bmatrix}
\boldsymbol{I}_{K}& \dots &\boldsymbol{ 0}_{K,K}&\boldsymbol{Z_{1N}}\boldsymbol{Q}(\boldsymbol{I}_K-\boldsymbol{Z_{NN}Q})^{-1}\\ 
\boldsymbol{0}_{K\times K} & \ddots &\vdots& \vdots\\ 
\vdots& \vdots & \boldsymbol{I}_{K}&\boldsymbol{Z_{(N-1)N}Q}(\boldsymbol{I}_K-\boldsymbol{Z_{NN}Q})^{-1}\\
\boldsymbol{0}_{K \times K}&\dots &\boldsymbol{0}_{K \times K}&(\boldsymbol{I}_K-\boldsymbol{Z_{NN}Q})^{-1}
\end{bmatrix}
\end{flalign}

\textbf{Step 5}: Finally, if we plug $\boldsymbol{X}^{-1}$ in Eq. (\ref{eq:X}), we will reach Eq. (\ref{eq:M}).\\

If we denote $K \times K$ diagonal blocks of $\boldsymbol{M}^{-1}$ as $[\boldsymbol{M}^{-1}]_{i,i}$ $1 \leq i \leq N$, then, by use of Lemma \ref{lemma:block}, trace($(\boldsymbol{R^{N+1}})^{-1}$) can be written as:
\begin{flalign}\label{eq:trace}
\nonumber
&\text{trace}((\boldsymbol{R^{N+1}})^{-1})=\sum_{i=1}^{N}{\text{trace}([\boldsymbol{M}^{-1}]_{i,i})}+\\
&+\text{trace}(\boldsymbol{R_{11}}^{-1}+\boldsymbol{R_{11}}^{-1}\boldsymbol{R_{21}}[\boldsymbol{M}^{-1}]_{N,N}\boldsymbol{R_{12}}\boldsymbol{R_{11}}^{-1})
\end{flalign}
By simplifying Eq. (\ref{eq:M}), diagonal blocks of $\boldsymbol{M}^{-1}$ can be presented as follows:
\begin{flalign}
\nonumber
&1\leq i \leq N-1:\\
\label{eq:n-1}
&[\boldsymbol{M}^{-1}]_{i,i}=\boldsymbol{Z_{ii}}+\boldsymbol{Z_{iN}Q}(\boldsymbol{I}-\boldsymbol{Z_{NN}Q})^{-1}\boldsymbol{Z_{Ni}} \\
\nonumber
& i=N:\\
\label{eq:n}
&[\boldsymbol{M}^{-1}]_{i,i}=(\boldsymbol{I}-\boldsymbol{Z_{NN}Q})^{-1}\boldsymbol{Z_{NN}} 
\end{flalign} 
In Eq. (\ref{eq:trace}), we set the diagonal blocks of $\boldsymbol{M}^{-1}$ as Eqs. (\ref{eq:n-1}) and (\ref{eq:n}). Then, by some manipulations, trace$((\boldsymbol{R^{N+1}})^{-1})$ can be presented as:
\begin{flalign}
\nonumber
&\text{trace}((\boldsymbol{R^{N+1}})^{-1})=\text{trace}((\boldsymbol{R^{N}})^{-1})+\text{trace}(\boldsymbol{R_{11}}^{-1})\\
\nonumber
&+\sum_{i=1}^{N-1}{\text{trace}(\boldsymbol{Z_{iN}Q}(\boldsymbol{I}-\boldsymbol{Z_{NN}Q})^{-1}\boldsymbol{Z_{Ni}})}\\
\nonumber
&+\text{trace}((\boldsymbol{I}-\boldsymbol{Z_{NN}Q})^{-1}\boldsymbol{Z_{NN}})-\text{trace}(\boldsymbol{Z_{NN}})\\
\label{eq:trace2}
&+\text{trace}(\boldsymbol{R_{11}}^{-1}\boldsymbol{R_{21}}(\boldsymbol{I}-\boldsymbol{Z_{NN}Q})^{-1}\boldsymbol{Z_{NN}R_{12}R_{11}}^{-1})
\end{flalign}
The first and second terms in Eq. (\ref{eq:trace2}) can be calculated by induction hypothesis and induction base, respectively. Calculating other terms in Eq. (\ref{eq:trace2}) is tedious but similar for different values of $K$. Therefore, we only calculate it for $K=2$ and skip the rest. For $K=2$, $\boldsymbol{Q}$ is equal to:
\begin{flalign}
\boldsymbol{Q}=\begin{bmatrix}
0& 0\\ 
0& \frac{1-\tau}{1+\tau}
\end{bmatrix}
\end{flalign}
where $\tau=\tau_2-\tau_1$.\\
If we plug $\boldsymbol{Q}=\begin{bmatrix}
0& 0\\ 
0& \frac{1-\tau}{1+\tau}
\end{bmatrix}$ in Eq. (\ref{eq:trace2}), after some calculations we will have:
\begin{flalign}\label{eq:two2}
\nonumber
&\text{trace}((\boldsymbol{R^{N+1}})^{-1})=\text{trace}((\boldsymbol{R^{N}})^{-1})+\frac{2}{1-(1-\tau)^2}\\
\nonumber
&+\frac{1-\tau}{(1+\tau)-(1-\tau)r(2N,2N)}\sum_{i=1}^{2N}{(r(2N,i))^2}\\
&+\frac{(1+\tau)(1+(\tau-1)^2)}{(2-\tau)^2[(1+\tau)-(1-\tau)r(2N,2N)]}r(2N,2N)
\end{flalign}
where $r(i,j)$ is the $(i,j)$th element of matrix $(\boldsymbol{R^N})^{-1}$. By induction hypothesis, the first term in Eq. (\ref{eq:two2}) is equal to $\frac{(N-1)(N+1)}{3(1-\tau_2+\tau_1)}+\frac{N(N+2)}{3(\tau_2-\tau_1)}+\frac{2N+1}{3(N+1-\tau_2+\tau_1)}$. For calculating Eq. (\ref{eq:two2}), we also need values of $r(2N,i)$, $1 \leq i \leq 2N$, which are elements of the last row of $(\boldsymbol{R^N})^{-1}$. Due to the special structure of matrix $\boldsymbol{R}$, values of $r(2N,i)$ can be calculated as follows:
\begin{flalign}\label{eq:pattern}
\left\{\begin{matrix}
r(2N,2i-1)=\frac{\tau-i}{\tau(N+1-\tau)} \\ 
r(2N,2i)=\frac{i}{\tau(N+1-\tau)}
\end{matrix}\right. 1\leq i\leq N
\end{flalign}
To verify Eq. (\ref{eq:pattern}), we can multiply the last row of $(\boldsymbol{R^N})^{-1}$, i.e., $[r(2N,1),r(2N,2),\dots,r(2N,2N)]$, by different columns of $\boldsymbol{R^N}$ as follows:
\begin{flalign}
\nonumber
&1\text{st column:}\ \ \frac{(\tau-1)1}{\tau(N+1-\tau)}+\frac{(1)(1-\tau)}{\tau(N+1-\tau)}=0\\
\nonumber
&\text{$(2i)_{th}$ column:} \ \ \ 1\leq i\leq N-1 \\
\nonumber
& \frac{(\tau-i)(1-\tau)}{\tau(N+1-\tau)}+ \frac{(i)1}{\tau(N+1-\tau)}+\frac{(\tau-(i+1))\tau}{\tau(N+1-\tau)}=0 \\
\nonumber
&\text{$(2i-1)_{th}$ column:} \ \ \ 2\leq i\leq N \\
\nonumber
& \frac{((i-1))(\tau)}{\tau(N+1-\tau)}+ \frac{(\tau-i)1}{\tau(N+1-\tau)}+\frac{(i)(1-\tau)}{\tau(N+1-\tau)}=0\\
\nonumber
&2N\text{th column:}\ \ \frac{(\tau-N)(1-\tau)}{\tau(N+1-\tau)}+\frac{(N)1}{\tau(N+1-\tau)}=1
\end{flalign}
These results verify that the last row of $(\boldsymbol{R^N})^{-1}$ follows the pattern in Eq. (\ref{eq:pattern}).\\
The last step is to plug Eq. (\ref{eq:pattern}) into Eq. (\ref{eq:two2}). As a result, trace($(\boldsymbol{R^{N+1})^{-1}}$) is equal to $\frac{(N)(N+2)}{3(1-\tau_2+\tau_1)}+\frac{(N+1)(N+3)}{3(\tau_2-\tau_1)}+\frac{2N+3}{3(N+2-\tau_2+\tau_1)}$, which verifies the induction step and completes the proof. 

\end{document}